\documentstyle[11pt,fleqn]{article}
\begin{document}

\begin{titlepage}

\begin{flushright}
Freiburg--THEP 94/05\\
May 1994
\end{flushright}
\vspace{1.5cm}

\begin{center}
\large\bf
{\LARGE\bf Massive two--loop diagrams:}\\
{\LARGE\bf The Higgs propagator}\\[1cm]
\rm
{A. Ghinculov}\\
{and}\\
{J.J. van der Bij}\\[.5cm]

{\em Albert--Ludwigs--Universit\"{a}t Freiburg,
           Fakult\"{a}t f\"{u}r Physik}\\
      {\em Hermann--Herder Str.3, D-79104 Freiburg, Germany}\\[1.5cm]

\end{center}
\normalsize

\begin{abstract}

Formulae are provided to express any two-loop scalar
integral with arbitrary masses and arbitrary
external momenta in terms of an integral
of one fairly simple function or of its derivatives.
Such integrals can be calculated numerically
with high precision. Good agreement is found with
known analytical expressions of specific
two--loop diagrams. To prove the effectiveness
of these techniques, the two--loop selfenergy
of the Higgs boson to order
$(g^{2} \frac{m_{H}^{2}}{m_{W}^{2}})^{2}$
is calculated, and the correctness of the result is checked
by using the unitarity of the S matrix.
This result allows one to disentangle the
leading corrections to the shape
of the Higgs resonance, and to derive
a perturbative bound of $\sim$1.2 TeV
on the mass of the Higgs boson.
\end{abstract}

\vspace{3cm}

\end{titlepage}


\section{Introduction}

Perturbation theory is practically the only known way
to compare field theory to the experiments.
Nonperturbative approaches, like lattice field theory, are
still far from reaching the precision level
of perturbation theory or of the LEP experiments.
The need for techniques to reliably calculate higher order
radiative corrections is therefore hard to overemphasize.

In the case of QCD, two--, three-- and even four--loop calculations
were performed \cite{larin,timo},
but the massive case is much more difficult,
and an analytical solution of the general two--loop integral
does not exist.

If the particles on the external lines are light, so that one
can perform an expansion around zero in the external momenta,
the Feynman integrals can be carried out analytically.
Such closed formulae made it possible for instance to calculate
certain two--loop electroweak corrections to the $\rho$ parameter
\cite{vdBij:2loop:rho}, to the masses of the vector bosons
\cite{vdBij:2loop:masses}, and to the selfcouplings of the vector bosons
\cite{vdBij:2loop:vertex}.

There are however physically interesting cases when such
an expansion cannot be performed. One obvious example is
the Higgs physics. Knowledge of higher order quantum corrections
in the Higgs sector  is interesting for several reasons.
First, if the Higgs particle is heavy, they will be numerically
important and will play a r\^{o}le in the Higgs searches at future colliders.
Second, as the Higgs mass increases, the Higgs
sector becomes strongly selfcoupled, and at some point
the perturbative expansion will break down. Knowledge of the
higher order corrections is the only way to find out up to which
point the calculations by
Feynman diagrams still can be trusted. Third,
having heavy particles on the external legs is the only way to
see the details of the symmetry breaking mechanism, which are
otherwise hidden by the screening theorem
\cite{veltman:screening,einhorn:screening}.

Considerable effort was devoted lately
to the calculation of massive two--loop Feynman diagrams.
Analytical results
exist for certain (mainly selfenergy) diagrams evaluated at
special values of the masses and of the external momenta.
Small and large momentum expansions were worked out, too
\cite{davydychev:small,davydychev:large},
although the small momentum expansion is
a rather trivial problem, since the propagators can be expanded
in the external momenta, and closed expressions for
the zero momentum case are available.
For the two--loop selfenergy diagram with
three propagators, an expression in terms
of generalized hypergeometric functions has been found
recently \cite{lauricella}.

These methods refer to special cases and were so far of little use
to calculate physically relevant quantities.

Recently a numerical approach to the so--called selfenergy
master diagram was proposed which reduces this diagram to
a two--fold integral \cite{kreimer}. It is possible to extend this method
to all selfenergy diagrams \cite{berends}. The generalization requires
a special treatment of different topologies, and also numerical
problems were reported in the imaginary part for certain values
of the masses and momenta \cite{berends}.

One would ideally want to use one single method
for any topology of two--loop diagrams. This would
allow one to write computer programs to generate and
compute automatically all Feynman diagrams needed for
a certain Green's function. Furthermore, one would like
to push the analytical integrations as far as possible
to increase the accuracy of the final result. The Monte--Carlo
techniques allow integrations over many variables, but their
accuracy is usually limited.

This paper proposes a method to reduce any two--loop massive
Feynman diagram to a standard form involving essentially
one basic function and its derivatives, by use of
the Feynman parameters. Further, this basic function
is integrated analytically as far as possible. Its
analyticity properties are discussed, leading to efficient
methods to perform the integration over the remaining
Feynman parameters numerically. Good agreement is found
with known analytical results. As an exercise,
we calculate in the last section
the selfenergy of the Higgs boson to order
$(g^{2} \frac{m_{H}^{2}}{m_{W}^{2}})^{2}$ ,
check the result by using the unitarity of the S matrix,
and extract the leading corrections to the
shape of the Higgs resonance.


\section{Reduction formulae}

In this section we provide formulae to reduce any two-loop
scalar diagram with arbitrary masses and arbitrary external
momenta to a standard form. This is an integral over
Feynman parameters of one basic function
or of its derivatives. In one special case (the two--loop integral
with three propagators), a second function needs to be introduced.
This second function appears rarely in practical calculations
and can in some cases be evaluated analytically.

The methods described in this section are easy to encode
into an algebraic computer program like FORM \cite{vermaseren},
in order to generate and handle
the large number of Feynman diagrams which may be required to
calculate Green's functions of physical relevance.

The most general form of a dimensionally regularized two--loop
scalar diagram in Minkowski space is:

\begin{eqnarray}
   I & = & \int d^{n}p\,d^{n}q\,
       \prod_{i=1}^{a} \frac{1}{(p+k_{i})^{2}-m_{i}^{2}+i \eta}\,
       \prod_{j=1}^{b} \frac{1}{(q+k_{a+j})^{2}-m_{a+j}^{2}+i \eta}
       \nonumber \\
  & &  \prod_{l=1}^{c} \frac{1}{(r+k_{a+b+l})^{2}-m_{a+b+l}^{2}+i \eta}
    \; \; ,
\end{eqnarray}
where $r=p+q$ is the sum of the loop momenta, and
$\{k_{i}\}_{i=1,\ldots,a+b+c}$ are external momenta, not
all of them independent.

As an obvious step, it is sometimes possible to reduce
the number of propagators by partial fractioning:

\begin{eqnarray}
\lefteqn{\frac{1}{[(p+k)^{2}-m_{1}^{2}]
                  [(p+k)^{2}-m_{2}^{2}]}  \, =}  \nonumber \\
& &      \frac{1}{m_{1}^{2}-m_{2}^{2}} \,
    [ \, \frac{1}{(p+k)^{2}-m_{1}^{2}} -
         \frac{1}{(q+k)^{2}-m_{2}^{2}}   \, ]
    \; \; .
\end{eqnarray}

One then combines all propagators with the same loop momentum
by using Feynman parameters:

\begin{eqnarray}
\lefteqn{\frac{1}{[(p+k_{1})^{2}-m_{1}^{2}]^{\alpha_{1}}
                  [(p+k_{2})^{2}-m_{2}^{2}]^{\alpha_{2}}}
		                             \, =}  \nonumber \\
& &      \frac{\Gamma(\alpha_{1}+\alpha_{2})}{\Gamma(\alpha_{1}) \,
	                                     \Gamma(\alpha_{2})}
        \int_{0}^{1} dx\,
        \frac{x^{\alpha_{1}-1}(1-x)^{\alpha_{2}-1}}{
	 [(p+\tilde{k})^{2}-\tilde{m^{2}}]^{\alpha_{1}+\alpha_{2}}}
    \; \; ,
\end{eqnarray}
where

\begin{eqnarray}
  \tilde{k}     & = & x k_{1} + (1-x) k_{2} \nonumber \\
  \tilde{m^{2}} & = & x m_{1}^{2} + (1-x) m_{2}^{2}
                  - x (1-x) (k_{1}-k_{2})^{2}
      \; \; .
\end{eqnarray}

The integral of eq. 1 can therefore be written in the following form:

\begin{eqnarray}
   I & = & \int dX\,d^{n}p\,d^{n}q\,
            P(x_{1},x_{2},\ldots,x_{a+b+c-3})
             \nonumber \\
     & &    \frac{1}{
             [(p+\tilde{k_{1}})^{2}-\tilde{m_{1}^{2}}]^{\alpha_{1}}
             [(q+\tilde{k_{2}})^{2}-\tilde{m_{2}^{2}}]^{\alpha_{2}}
             [(r+\tilde{k_{3}})^{2}-\tilde{m_{3}^{2}}]^{\alpha_{3}}
	    }
    \; \; ,
\end{eqnarray}
where P is some polynomial, $x_{1},x_{2},\ldots,x_{a+b+c-3}$ are
the Feynman parameters (we assume that no partial fractioning
was possible in eq. 1), and
$X\ = dx_{1}\,dx_{2}\,\ldots\,dx_{a+b+c-3}$.
$\tilde{k_{1}}$, $\tilde{k_{2}}$ and $\tilde{k_{3}}$
are polynomials of the external momenta and of the Feynman parameters,
and $\tilde{m_{1}^{2}}$, $\tilde{m_{2}^{2}}$ and $\tilde{m_{3}^{2}}$
are polynomials of the masses,
the Mandelstam variables, and the Feynman parameters.

To simplify the notations, we will drop in the following the tildes,
the polynomial P, and the integration over the Feynman parameters,
keeping in mind that the masses and the momenta
depend on $x_{1},x_{2},\ldots,x_{a+b+c-3}$.

By redefining the loop momenta, we can rewrite eq. 5 in the
following form:

\begin{equation}
   I = \int d^{n}p\,d^{n}q\,
       \frac{1}{
             (p^{2}-m_{1}^{2})^{\alpha_{1}}
             (q^{2}-m_{2}^{2})^{\alpha_{2}}
             [(r+k_{3}-k_{1}-k_{2})^{2}-m_{3}^{2}]^{\alpha_{3}}
	    }
    \; \; .
\end{equation}

The Lorentz invariance implies that $I$ depends only on
$(k_{3}-k_{1}-k_{2})^2$. It is also obvious that

\begin{eqnarray}
\int d^{n}p\,d^{n}q\,
       \frac{1}{
             [(p+k)^{2}-m_{1}^{2}]^{\alpha_{1}} \,
             (q^{2}-m_{2}^{2})^{\alpha_{2}} \,
             (r^{2}-m_{3}^{2})^{\alpha_{3}}
	    }   & = &  \nonumber \\
\int d^{n}p\,d^{n}q\,
       \frac{1}{
             (p^{2}-m_{1}^{2})^{\alpha_{1}} \,
             [(q+k)^{2}-m_{2}^{2}]^{\alpha_{2}} \,
             (r^{2}-m_{3}^{2})^{\alpha_{3}}
	    }   & = &  \nonumber \\
\int d^{n}p\,d^{n}q\,
       \frac{1}{
             (p^{2}-m_{1}^{2})^{\alpha_{1}} \,
             (q^{2}-m_{2}^{2})^{\alpha_{2}} \,
             [(r+k)^{2}-m_{3}^{2}]^{\alpha_{3}}
	    } & &
      \; \; .
\end{eqnarray}

These properties justify the notation:

\begin{eqnarray}
\lefteqn{G(m_{1},\alpha_{1};
           m_{2},\alpha_{2};
           m_{3},\alpha_{3};k^2)  \, =}  \nonumber \\
& &    \int d^{n}p\,d^{n}q\,
       \frac{1}{
             (p^{2}+m_{1}^{2})^{\alpha_{1}} \,
             (q^{2}+m_{2}^{2})^{\alpha_{2}} \,
             [(r+k)^{2}+m_{3}^{2}]^{\alpha_{3}}
	    }
    \; \; ,
\end{eqnarray}
where the momenta are now Euclidian.

We further notice that

\begin{eqnarray}
\lefteqn{G(m_{1},\alpha_{1}+1;
           m_{2},\alpha_{2};
           m_{3},\alpha_{3};k^2) \, =}  \nonumber \\
& &  - \frac{1}{\alpha_{1}} \frac{\partial}{\partial m_{1}^{2}}
        G(m_{1},\alpha_{1};
          m_{2},\alpha_{2};
          m_{3},\alpha_{3};k^2)
    \; \; .
\end{eqnarray}

Such relations allow one to express all higher order G functions
in terms of the derivatives of one basic function. We choose
this function to be
$G(m_{1},2;m_{2},1;m_{3},1;k^2)$ , and will denote it in the following
by
\linebreak
${\cal G}(m_{1},m_{2},m_{3};k^2)$.
The diagrammatical representation of ${\cal G}$
and of its derivatives is shown in Fig. 1.

In this way, one case remains uncovered, namely
$G(m_{1},1;m_{2},1;m_{3},1;k^2)$. We do not choose this function
as the basic function because of its bad ultraviolet properties.
Namely, if one na\"{\i}vely uses the Feynman parameters
to calculate it, part of the ultraviolet divergencies will
be transferred from the radial integration
to the integration over the Feynman parameters.

To calculate $G(m_{1},1;m_{2},1;m_{3},1;k^2)$,
we start by using the "partial p" operation \cite{tHooft}:

\begin{eqnarray}
\lefteqn{G(m_{1},1;m_{2},1;m_{3},1;k^2) \, \equiv}  \nonumber \\
& & \int d^{n}p\,d^{n}q\,
       \frac{1}{
             (p^{2}+m_{1}^{2})
             [(q+k)^{2}+m_{2}^{2}]
             (r^{2}+m_{3}^{2})
	    } \,\,  =  \nonumber \\
& & \frac{2}{n} \int d^{n}p\,d^{n}q\,
     \{  \frac{p^{2}}{
             (p^{2}+m_{1}^{2})^{2}
             [(q+k)^{2}+m_{2}^{2}]
             (r^{2}+m_{3}^{2}) }
	    \nonumber \\
& &    + \frac{p^{2}+p.q}{
             (p^{2}+m_{1}^{2})
             [(q+k)^{2}+m_{2}^{2}]
             (r^{2}+m_{3}^{2}) ^{2}  }
     \}
      \; \; .
\end{eqnarray}

After some algebra and explicit symmetrization upon
$m_{1} \leftrightarrow m_{2}$ , one ends up with
the following reduction formula:

\begin{eqnarray}
\lefteqn{G(m_{1},1;m_{2},1;m_{3},1;k^2) \, =}  \nonumber \\
& &  \frac{1}{3-n}
     \{   m_{1}^{2} \, {\cal G}(m_{1}, m_{2}, m_{3} ;k^2)
        + m_{2}^{2} \, {\cal G}(m_{2}, m_{1}, m_{3} ;k^2)
	                  \nonumber \\
& &  \; \; \; \;   + m_{3}^{2} \, {\cal G}(m_{3}, m_{1}, m_{2} ;k^2)
        +           {\cal F}(m_{1}, m_{2}, m_{3} ;k^2)
     \}
      \; \; ,
\end{eqnarray}
where

\begin{eqnarray}
\lefteqn{{\cal F}(m_{1}, m_{2}, m_{3} ;k^2) \, =}  \nonumber \\
& & - \int d^{n}p\,d^{n}q\,
       \frac{(p+q).k}{
             (p^{2}+m_{1}^{2})
             [(q+k)^{2}+m_{2}^{2}]
             (r^{2}+m_{3}^{2})^{2}
	    }
      \; \; .
\end{eqnarray}

The function ${\cal F}$ must be symmetrical under permuting
$m_{1}$, $m_{2}$ and $m_{3}$ due to the original symmetry
of $ G(m_{1},1;m_{2},1;m_{3},1;k^2)$ and to the
\mbox{$m_{2} \leftrightarrow m_{3}$} symmetry of
${\cal G}(m_{1}, m_{2}, m_{3} ;k^2)$. Another property of
${\cal F}$ is that it vanishes in the limit $k^2 \rightarrow 0$.
Indeed, the function ${\cal G}(m_{1}, m_{2}, m_{3} ;k^2)$
reduces in this limit to the symbol
$(m_{1},m_{1} | m_{2} | m_{3})$ of ref. \cite{vdBij:2loop:rho}.
According to ref. \cite{vdBij:2loop:rho}, the following formula
holds:

\begin{eqnarray}
\lefteqn{G(m_{1},1;m_{2},1;m_{3},1;0) \, =}  \nonumber \\
& &  \frac{1}{3-n}
     \{   m_{1}^{2} \, {\cal G}(m_{1}, m_{2}, m_{3} ;0)
        + m_{2}^{2} \, {\cal G}(m_{2}, m_{1}, m_{3} ;0)
	                               \nonumber \\
& &   \; \; \; \;  + m_{3}^{2} \, {\cal G}(m_{3}, m_{1}, m_{2} ;0)
     \}
      \; \; .
\end{eqnarray}
This proves the statement about the limit of ${\cal F}$.

We can at this point express any two--loop Feynman diagram
as an integral over ${\cal G}$ functions or over the
derivatives of ${\cal G}$. In the case of the two--loop diagram with
three propagators and nonvanishing external momentum,
one ${\cal F}$ function appears, too.


\section{Analytical integration}

In this section we integrate the ${\cal F}$ and ${\cal G}$
functions analytically as far as possible. One will be left
in the general case with an integral over one Feynman parameter.
Only in special cases can this last integral be solved
analytically. However, a careful examination of the integrand's
analyticity properties allows one to write fast computer programs
to evaluate these functions numerically with high precision.

We start with the ${\cal G}$ function. After introducing
two Feynman parameters and integrating over the loop variables,
one obtains:

\begin{eqnarray}
\lefteqn{{\cal G}(m_{1},m_{2},m_{3};k^2) \, \equiv}  \nonumber \\
& & \int d^{n}p\,d^{n}q\,
       \frac{1}{
             (p^{2}+m_{1}^{2})^{2} \,
             [(q+k)^{2}+m_{2}^{2}] \,
             [(p+q)^{2}+m_{3}^{2}]
	    } \,  =  \nonumber \\
& & \pi^{4} (\pi \, m_{1}^{2})^{n-4}
    \frac{\Gamma(2-\frac{n}{2})}{\Gamma(3-\frac{n}{2})}
    \int_{0}^{1} dx\, \int_{0}^{1} dy\,
    [x (1-x)]^{\frac{n}{2}-2} y (1-y)^{2-\frac{n}{2}}
                                                \nonumber \\
& & \{ \, \Gamma(5-n) \frac{y^{2} \kappa^{2} + \mu^{2}}{[y(1-y)\kappa^{2}
                        + y + (1-y)\mu^{2}]^{5-n}}
                                                \nonumber \\
& &   + \frac{n}{2} \Gamma(4-n) \frac{1}{[y(1-y)\kappa^{2}
                        + y + (1-y)\mu^{2}]^{4-n}}
     \, \}
      \; \; ,
\end{eqnarray}
where the following notations were introduced:

\begin{eqnarray}
   \mu^{2}  & = &  \frac{a x + b (1-x)}{x (1-x)}   \nonumber \\
         a  & = &  \frac{m_{2}^{2}}{m_{1}^{2}} \, , \; \; \; \;
         b \; = \; \frac{m_{3}^{2}}{m_{1}^{2}} \, , \; \; \; \;
\kappa^{2} \; = \; \frac{    k^{2}}{m_{1}^{2}}
      \; \; .
\end{eqnarray}

We then expand in $\epsilon = n - 4$ and integrate over $y$ .
After some tedious but elementary algebra, one obtains:

\begin{eqnarray}
\lefteqn{{\cal G}(m_{1},m_{2},m_{3};k^2) \, =}  \nonumber \\
& &   \pi^{4} \{ \,   \frac{2}{\epsilon^{2}}
    + \frac{1}{\epsilon} [- 1 + 2 \gamma + 2 \log (\pi \, m_{1}^{2}) ]
    + \frac{1}{4} + \frac{\pi^{2}}{12}
                                                \nonumber \\
& & + \frac{1}{4} [- 1 + 2 \gamma + 2 \log (\pi \, m_{1}^{2}) ]^{2}
    - 1 + g(m_{1},m_{2},m_{3};k^2)
          \,    \}  \; \; ,
\end{eqnarray}
where

\begin{eqnarray}
g(m_{1},m_{2},m_{3};k^2) & =&   \int_{0}^{1}\,dx\,
     [ \, Sp(\frac{1}{1-y_{1}}) + Sp(\frac{1}{1-y_{2}})
                                                \nonumber \\
& & + y_{1}\log \frac{y_{1}}{y_{1}-1} +
      y_{2}\log \frac{y_{2}}{y_{2}-1} \, ]
      \; \; ,
\end{eqnarray}
and

\begin{eqnarray}
y_{1,2} & = & \frac{1 + \kappa^{2} - \mu^{2}
                    \pm \sqrt{\Delta}}{2 \kappa^{2}}  \nonumber \\
\Delta  & = & (1 + \kappa^{2} - \mu^{2})^{2}
          + 4 \kappa^{2} \mu^{2} - 4 i \kappa^{2} \eta
      \; \; .
\end{eqnarray}

In general the finite part $g$ cannot be integrated analytically.
However, this is possible in the trivial case
when $k^{2} \rightarrow 0$ . One can convince oneself that eq. 17
reduces in this limit to

\begin{equation}
g(m_{1},m_{2},m_{3};0) = \int_{0}^{1} dx \,
 [ \, 1 + Sp(1-\mu^{2}) - \frac{\mu^{2}}{1-\mu^{2}} \log \mu^{2} \, ]
      \; \; ,
\end{equation}
which is essentially the finite part of the function
$(m_{1},m_{1}|m_{2}|m_{3})$ of ref. \cite{vdBij:2loop:rho}.

In this case the $x$ integration can be carried out, and gives:

\begin{eqnarray}
\lefteqn{g(m_{1},m_{2},m_{3};0) \, =}         \nonumber \\
& & 1 - \frac{1}{2} \log a  \log b  - \frac{a+b-1}{\sqrt{\Delta \prime}}
     [ \,  Sp(-\frac{u_{2}}{v_{1}}) + Sp(-\frac{v_{2}}{u_{1}})
                                                \nonumber \\
& &   + \frac{1}{4} \log^{2} \frac{u_{2}}{v_{1}}
      + \frac{1}{4} \log^{2} \frac{v_{2}}{u_{1}}
      + \frac{1}{4} \log^{2} \frac{u_{1}}{v_{1}}
      - \frac{1}{4} \log^{2} \frac{u_{2}}{v_{2}}
      + \frac{\pi^{2}}{6}
   \,  ]
      \; \; ,
\end{eqnarray}
where

\begin{eqnarray}
u_{1,2}        & = & \frac{1}{2}
              ( 1 + b - a
                    \pm \sqrt{\Delta^{\prime}} )  \nonumber \\
v_{1,2}        & = & \frac{1}{2}
              ( 1 - b + a
                    \pm \sqrt{\Delta^{\prime}} )  \nonumber \\
\Delta^{\prime}  & = & 1 - 2 (a+b) + (a-b)^{2}
      \; \; .
\end{eqnarray}

This agrees with the results of ref. \cite{vdBij:2loop:rho} .
The result of eq. 20 implies that any massive two--loop Feynman diagram
can be calculated analytically if all Mandelstam variables vanish.

Let us now turn to the ${\cal F}$ function. Eq. 12
can be rewritten in the following form:

\begin{displaymath}
{\cal F}(m_{1},m_{2},m_{3};k^2) =
k^{2} {\cal G}(m_{1},m_{2},m_{3};k^2) +
{\cal F}^{\prime}(m_{1},m_{2},m_{3};k^2)
\end{displaymath}
\begin{eqnarray}
\lefteqn{{\cal F}^{\prime}(m_{1},m_{2},m_{3};k^2) \, =}
                                                \nonumber \\
& & \int d^{n}p\,d^{n}q\,
       \frac{p.k}{
             [(p+k)^{2}+m_{3}^{2}]^{2} \,
             (q^{2}+m_{1}^{2}) \,
             [(p+q)^{2}+m_{2}^{2}] }
      \; \; .
\end{eqnarray}

We combine the propagators by using Feynman parameters,
and integrate over the loop momenta:

\begin{eqnarray}
\lefteqn{{\cal F}^{\prime}(m_{1},m_{2},m_{3};k^2) \, =}
                                                \nonumber \\
& & - k^{2} \pi^{4} (\pi m_{1}^{2})^{n-4}
    \frac{\Gamma(2-\frac{n}{2})}{\Gamma(3-\frac{n}{2})}
    \int_{0}^{1} dx\, \int_{0}^{1} dy\,
    [x (1-x)]^{\frac{n}{2}-2} y^{2} (1-y)^{2-\frac{n}{2}}
                                                \nonumber \\
& & \{ \Gamma(5-n) \frac{y^{2} \kappa^{2} + \mu^{2}}{[y(1-y)\kappa^{2}
                        + y + (1-y)\mu^{2}]^{5-n}}
                                                \nonumber \\
& &   + \frac{n}{2} \Gamma(4-n) \frac{1+\frac{2}{n}}{[y(1-y)\kappa^{2}
                        + y + (1-y)\mu^{2}]^{4-n}}
    \}
      \; \; ,
\end{eqnarray}
with the notations of eq. 15.

One next expands in $\epsilon = n - 4$ and integrates $y$ out.
Using the result for ${\cal G}(m_{1},m_{2},m_{3};k^2)$ ,
one finds:

\begin{eqnarray}
\lefteqn{{\cal F}(m_{1},m_{2},m_{3};k^2) \, =}  \nonumber \\
& & k^{2} \pi^{4} \{ \, - \frac{1}{2 \epsilon}
                     + \frac{9}{8}
    - \frac{1}{2} [ \gamma + \log (\pi m_{1}^{2}) ]
    + f(m_{1},m_{2},m_{3};k^2)
          \,    \}  \; \; ,
\end{eqnarray}
where

\begin{eqnarray}
\lefteqn{f(m_{1},m_{2},m_{3};k^2) \, =}         \nonumber \\
& & \int_{0}^{1}\,dx\,
     [ \, \frac{1-\mu^{2}}{2 \kappa^{2}}
    - \frac{1}{2} \, y_{1}^{2} \, \log \frac{y_{1}}{y_{1}-1}
    - \frac{1}{2} \, y_{2}^{2} \, \log \frac{y_{2}}{y_{2}-1} \,  ]
      \; \; .
\end{eqnarray}

As already mentioned, the ${\cal F}$ function must be
invariant under permutations of the masses, and has to vanish
in the $k^{2} \rightarrow 0$ limit. While the
$m_{2} \leftrightarrow m_{3}$ symmetry is obvious,
the $m_{1} \leftrightarrow m_{2}$ symmetry is less trivial,
and was checked numerically.

The $f$ function is needed only for the two--loop diagrams
with three propagators, and can be integrated analytically for
special choices of the arguments.


\section{Analyticity and numerical integration}

We found that the finite part of the basic function ${\cal G}$
is expressed in the general case through a one dimensional
integral:

\begin{eqnarray}
g(m_{1},m_{2},m_{3};k^2) & =&   \int_{0}^{1}\,dx\,
\tilde{g}(m_{1},m_{2},m_{3};k^2;x)
                                                \nonumber \\
\tilde{g}(m_{1},m_{2},m_{3};k^2;x)
& = &   Sp(\frac{1}{1-y_{1}}) + Sp(\frac{1}{1-y_{2}}) +
                                                \nonumber \\
& & y_{1}\log \frac{y_{1}}{y_{1}-1} +
    y_{2}\log \frac{y_{2}}{y_{2}-1}
      \; \; ,
                                                \nonumber
\end{eqnarray}
\begin{eqnarray}
y_{1,2} & = & \frac{1 + \kappa^{2} - \mu^{2}
                    \pm \sqrt{\Delta}}{2 \kappa^{2}}  \nonumber \\
\Delta  & = & (1 + \kappa^{2} - \mu^{2})^{2}
          + 4 \kappa^{2} \mu^{2} - 4 i \kappa^{2} \eta
      \; \; .
\end{eqnarray}

It is straightforward to calculate the
derivatives of $g$ with respect to $m_{1}$, $m_{2}$ and $m_{3}$,
which are needed to evaluate Feynman diagrams
with more than four propagators.

In eq. 26, $k^{2} > 0$ corresponds to Euclidian momenta. When one rotates
$k$ back to Minkowski space to calculate the physical Green's
functions, the function $g$ and its derivatives
develop an imaginary part above the threshold
$-k^{2} > (m_{1}+m_{2}+m_{3})^{2}$ .
This is readily seen from the diagrammatical representation
of fig. 1.

The typical behaviour of $\tilde{g}$ as a function of the
integration variable $x$ is shown in fig.2.
It has singularities of the logarithmic type at $x = 0$ and $x = 1$ .
Once they are extracted through a suitable change of variable,
the function $g$ can easily be integrated numerically.

Problems start to appear when trying to evaluate Feynman diagrams
with more than 5 propagators, above the threshold.
For the case of six propagators there are two Dirac $\delta$ functions
appearing at the points where the derivative of the function $\tilde{g}$
is discontinuous, and starting with seven propagators one deals already
with the derivatives of the $\delta$ distribution.

To circumvent this kind of problems, one has to analytically
continue the $\tilde{g}$ function in the complex plane of the
Feynman parameter $x$, and to choose an integration path which
avoids the singularities.

The singular points of $\tilde{g}$, other than $x=0$ and $x=1$,
are given by the roots
of the discriminant in eq. 26. The equation $\Delta = 0$
has four roots:

\begin{eqnarray}
x_{1,2} & = & \frac{1}{2 \mu_{1}^{2}}
             \, [ \, -a + b + \mu_{1}^{2}
   \pm \sqrt{(a-b-\mu_{1}^{2})^{2} - 4 b \mu_{1}^{2}} \, ]  \nonumber \\
x_{3,4} & = & \frac{1}{2 \mu_{2}^{2}}
             \, [ \, -a + b + \mu_{2}^{2}
   \pm \sqrt{(a-b-\mu_{2}^{2})^{2} - 4 b \mu_{2}^{2}} \, ]
      \; \; ,
                                                         \nonumber \\
\mu_{1,2}^{2}  & = & 1 - \kappa^{2} \mp 2 \sqrt{- \kappa^{2}}
      \; \; .
\end{eqnarray}

These are branching points for each separate
term in the expression 26 of $\tilde{g}$ .
However, the $\tilde{g}$ function itself has only two branching points
at $x_{1}$ and $x_{2}$,
because the singularities
at $x_{3}$ and $x_{4}$ are compensating
among the four terms of $\tilde{g}$ in eq. 26.

This behaviour is related to the causality of the Green's functions.
The causality condition is expressed by the $i \eta$ prescription
in the Feynman propagator, wich means to replace all masses
$m^{2}$ by $m^{2}-i\eta$. An equivalent way to impose causality
is to calculate the Euclidian Green's functions and to go afterwards
to physical momenta, approaching the cut on the positive real axis
from above. This amounts to making the replacement
$k^{2} \rightarrow -k^{2}-i\eta$ in eq. 26.
These two prescriptions ought to be equivalent,
and therefore have to fix the location of the physical singularities
with respect to the real axis in the same way.
For $x_{1}$ and $x_{2}$, at $-\kappa^{2} > 1$
both prescriptions lead to the same change, and therefore these
are the singularities of the $\tilde{g}$ function.
For $x_{3}$ and $x_{4}$, the two prescriptions would lead to opposite
changes in the imaginary direction. Since causality fixes the location
of the singularities of the Green's function uniquely,
$x_{3}$ and $x_{4}$ cannot correspond to real singularities of $\tilde{g}$.
Therefore $\tilde{g}$ is analytical at these two points.

Fig. 3 shows the typical behaviour of $\tilde{g}$ above the
threshold as a function of the complex Feynman parameter $x$ .
One can now choose a complex integration path on the physical
Riemann sheet along which the integrand is free of singularities.
On such a path the $\tilde{g}$ function is analytical,
and so are its derivatives of any order, which are requested
to calculate diagrams with more than four propagators.

To calculate Feynman diagrams, one needs to integrate the $g$
function or the derivatives of $g$ over the Feynman parameters
which were introduced to combine the propagators with the same
loop momentum. In general the arguments of $g$ will take negative
values, too, corresponding to imaginary masses in eq. 26. This
corresponds to various thresholds of the diagram. The points
where the mass arguments vanish must also be avoided, since $g$
and its derivatives may have mass singularities at these points.
One needs to make sure the integrand remains always on the
physical sheet. Therefore one always chooses the integration path
according to the Feynman prescription for the masses of the propagators:
$m^{2} \rightarrow m^{2} - i \eta$ .

A computer program based on this method was written
and checked upon known analytical and numerical results
for specific diagrams.

The topology of fig. 5 a) in the case
$s = 1$, $m_{1}^{2}=m_{2}^{2}= \ldots =m_{5}^{2} \rightarrow 0$ ,
which is related to the large momentum limit, was calculated
and found to agree with the known value of
$6 \zeta(3)$ \cite{tkachov}. Also the case
$s = 1$, $m_{1}^{2}=m_{3}^{2}=m_{4}^{2}=1$,
$m_{2}^{2}=m_{5}^{2} \rightarrow 0$ agrees with the analytical
result $\pi^{2} \log 2 - \frac{3}{2}\zeta(3)$ \cite{kreimer}.

For the diagram of fig. 5 h), an analytical formula in terms
of Lauricella functions was recently derived \cite{lauricella}.
The authors of ref. \cite{lauricella} calculate an ultraviolet
finite combination $T_{123N}$ of four such diagrams.
$T_{123N}$ can be expressed in terms of 12 $g$ functions
and 4 $f$ functions. Perfect agreement was found with
the numerical values which are given in ref. \cite{lauricella}
for a range of masses and momenta.

Finally, some comments on the numerical integration are in order.
The $g$ function can easily be integrated through virtually
any algorithm after extracting the singularities at the ends
of the integration path. These are of logarithmic type, and can
be extracted minimally through a change of variables of the type:

\begin{displaymath}
t = x( 24 - 24 \log x + 12 \log^{2} x - 4 \log^{3} x + \log^{4} x )
      \; \; .
\end{displaymath}

A simpler nonminimal change of variable, like $t=\sqrt{x}$ , can
be more handy. It puts the integrand to zero at the ends
of the integration path.

Because the integrand is smooth along the integration path
and free of violent variations, the $g$ function and its
derivatives can be integrated numerically very fast, using a small
number of points. Typically some 120 points were requested to calculate
these functions to more than 8 digits. This takes about 50 ms on
an IBM RISC 6000 workstation. Integrating the $g$ function or its
derivatives further in order to calculate Feynman diagrams with
more propagators poses also no problem as long as the appropriate complex
integration path is used in order to avoid the singularities.
This yields also a smooth function which can be integrated
with high precision. For less than four Feynman parameters, an
adaptative deterministic algorithm was used, yielding accurate results
already with a small number of points. For more than four dimensions the
Monte--Carlo techniques become superior, but one cannot hope to obtain
an accuracy better than 3--4 digits.


\section{The selfenergy of the Higgs boson}

To show how the techniques of the previous sections work,
and to prove that they can be used to perform reliable
and accurate calculations of physical relevance, we calculate
in this section the selfenergy of the Higgs boson at order
$(g^{2} \frac{m_{H}^{2}}{m_{W}^{2}})^{2}$.
We first calculate the on--shell selfenergy,
check the correctness of the results by using the
unitarity of the S matrix, and extract
the mass counterterm.
We then calculate the momentum dependence of the
Higgs selfenergy to see its analytic structure,
and check some of its asymptotic properties.
The result allows one to extract the leading
corrections to the shape of the Higgs resonance.
This is an effect which does not appear in the
selfenergy of the Higgs boson at one--loop level.
The leading corrections to the Higgs shape become large
if the Higgs mass is of the order of 1.2 TeV,
indicating a strongly interacting theory,
in agreement with well--known Born level and
one--loop results \cite{lee-quigg,marciano}.

Since we are interested in the leading contribution in $m_{H}$,
the most natural choice
is to work in Landau gauge. In this gauge only the Higgs sector
survives at leading order in $m_{H}$. The diagrams containing
gauge, fermion, or
Fadeev--Popov fields do not give contributions of
order $(g^{2} \frac{m_{H}^{2}}{m_{W}^{2}})^{2}$
in this gauge.

However, to avoid problems with massless Feynman
diagrams which can be traced back to the arbitrariness
of $\int d^{n}p \, \frac{1}{p^{4}}$
within the dimensional regularization, we choose to keep
a small gauge parameter $\xi$ during the computation,
and take the limit $\xi \rightarrow 0$ in the final result.
This amounts to keeping a finite mass of the
Goldstone bosons. The diagrams involving gauge and Fadeev--Popov
fields can safely be omitted, since they do not
generate finite terms in the limit $\xi \rightarrow 0$.
This provides also a useful check
of the computation, since  all poles and logarithms
of the gauge parameter have to cancel in the final result.

The Lagrangian of the Higgs--Goldstone system is:

\begin{eqnarray}
{\cal L} & = &
\frac{1}{2} (\partial_{\mu}H_{0})(\partial^{\mu}H_{0}) +
\frac{1}{2} (\partial_{\mu}z_{0})(\partial^{\mu}z_{0}) +
            (\partial_{\mu}w_{0}^{+})(\partial^{\mu}w_{0}^{-})
                                                \nonumber \\
& & - g^{2}\frac{m_{H_{0}}^{2}}{m_{W_{0}}^{2}} \frac{1}{8} \,
[ \, w_{0}^{+} w_{0}^{-} + \frac{1}{2} z_{0}^{2} + \frac{1}{2} H_{0}^{2}
+ \frac{2 m_{W_{0}}}{g} H_{0}
+ \frac{4 \, \delta t}{g^{2} \, \frac{m_{H_{0}}^{2}}{m_{W_{0}}^{2}}}
 \, ]^{2}
      \; \; ,
\end{eqnarray}
with obvious notations. $\delta t$ is the tadpole counterterm
needed to ensure that $<0|H|0> = v$ at all orders.

The renormalization is performed by splitting the bare quantities
in renormalized quantities and counterterms:

\begin{eqnarray}
H_{0} & = & Z_{H}^{1/2} H      \nonumber \\
z_{0} & = & Z_{G}^{1/2} z      \nonumber \\
w_{0} & = & Z_{G}^{1/2} w      \nonumber \\
m_{H_{0}}^{2} & = & m_{H}^{2} - \delta m_{H}^{2}     \nonumber \\
m_{W_{0}}^{2} & = & m_{W}^{2} - \delta m_{W}^{2}
      \; \; .
\end{eqnarray}

Note that the field renormalization constants of the charged
and the neutral Goldstone bosons can be chosen equal due to
the remnant $O(3)$ symmetry of the Lagrangian.

We adopt the tadpole renormalization strategy described
by Taylor \cite{taylor}. The condition that the vacuum
expectation value of the Higgs field should not receive
quantum corrections amounts to neglecting the tadpole
insertions altogether and subtracting the value of the
Goldstone bosons selfenergy at zero
momentum from all scalar selfenergies.

The counterterms are fixed uniquely through the
renormalization conditions. We use an on--shell renormalization
scheme with field renormalization, and the physical masses
of the Higgs and of the W bosons as input parameters. Note
that the gauge coupling constant does not get renormalized
at leading order in $g^{2} \frac{m_{H}^{2}}{m_{W}^{2}}$.

By evaluating the one--loop selfenergies of the Higgs, Goldstone
and W bosons, one finds the following renormalization constants
at ${\cal O}(g^{2} \frac{m_{H}^{2}}{m_{W}^{2}})$:

\begin{eqnarray}
\delta t^{(1-loop)} & = &  g^{2} \frac{m_{H}^{2}}{m_{W}^{2}} \,
                           (\frac{m_{H}^{2}}{4 \pi \mu^{2}})^{\epsilon /2} \,
                           \frac{m_{H}^{2}}{16 \pi^{2}} \, \{
- {{3}\over {4\,\epsilon }} + {{3}\over 8}  -
  {{3\, \gamma }\over 8}
     \nonumber \\
& &  + \epsilon \,\left( {{-3}\over {16}} +
     {{3\, \gamma }\over {16}} -
     {{3\,{{ \gamma }^2}}\over {32}} -
     {{{{\pi }^2}}\over {64}} \right)
  \}
     \nonumber \\
\delta m_{H}^{2\,(1-loop)} & = & g^{2} \frac{m_{H}^{2}}{m_{W}^{2}} \,
                           (\frac{m_{H}^{2}}{4 \pi \mu^{2}})^{\epsilon /2} \,
                           \frac{m_{H}^{2}}{16 \pi^{2}} \, \{ \,
{{3}\over {\epsilon }} - 3 +
  {{3\, \gamma }\over 2} +
  {{{3\,\sqrt{3}}\,\pi }\over 8}
     \nonumber \\
& &  + \epsilon \,[ 3 -
     {{3\, \gamma }\over 2} +
     {{3\,{{ \gamma }^2}}\over 8} -
     {{{3\,\sqrt{3}}\,\pi }\over 8} +
     {{{3\,\sqrt{3}}\, \gamma \,
         \pi }\over {16}}
     \nonumber \\
& &  - {{{{\pi }^2}}\over {16}} -
     {{{3\,\sqrt{3}}\,
         {\it Cl}({{\pi }\over 3})}\over 4} +
     {{{3\,\sqrt{3}}\,\pi \,\log (3)}\over
       {16}}  \, ]
  \}
     \nonumber \\
\delta m_{W}^{2\,(1-loop)} & = & g^{2} \frac{m_{H}^{2}}{m_{W}^{2}} \,
                           (\frac{m_{H}^{2}}{4 \pi \mu^{2}})^{\epsilon /2} \,
                           \frac{m_{W}^{2}}{16 \pi^{2}} \, [ \,
{{{ 1 }}\over 8} +
  \epsilon \,\left( - {{3}\over {32}} +
     {{ \gamma }\over {16}} \right)
 \, ]
     \nonumber \\
\delta Z_{H}^{(1-loop)} & = &  g^{2} \frac{m_{H}^{2}}{m_{W}^{2}} \,
                           (\frac{m_{H}^{2}}{4 \pi \mu^{2}})^{\epsilon /2} \,
                           \frac{1}{16 \pi^{2}} \, \{ \,
{3\over 2} -
    {{\pi \,{\sqrt{3}}\over 4}} + \epsilon \,
   [ \, -{3\over 2} + {{3\, \gamma }\over 4}
        \nonumber \\
& & + {{{3\,\sqrt{3}}\,\pi }\over {16}} -
    {{{\sqrt{3}}\, \gamma \,\pi }\over 8} +
     {{{\sqrt{3}}\,{\it Cl}({{\pi }\over 3})}\over 2} -
     {{{\sqrt{3}}\,\pi \,\log (3)}\over 8} \, ]
  \}
     \nonumber \\
\delta Z_{G}^{(1-loop)} & = &  g^{2} \frac{m_{H}^{2}}{m_{W}^{2}} \,
                           (\frac{m_{H}^{2}}{4 \pi \mu^{2}})^{\epsilon /2} \,
                           \frac{1}{16 \pi^{2}} \, [ \,
-{1\over 8} + \epsilon \,
   \left( {3\over {32}} -
     {{ \gamma }\over {16}} \right)
 \, ]
      \; \; .
\end{eqnarray}

One needs the one--loop counterterms at ${\cal O}(\epsilon)$
because they combine with the $\frac{1}{\epsilon}$
terms at two--loop order to give finite contributions.

We can now turn to the actual two--loop calculation.

The two--loop tadpole counterterm can be calculated from the diagrams of
fig. 4.
Alternatively, one can calculate the $w$ or $z$ selfenergies at zero momentum.
In any case, the tadpole counterterm can be evaluated analytically,
since it involves only the $g(m_{1},m_{2},m_{3};0)$ function and its
derivatives. It was checked that all three ways to calculate
the tadpole counterterm lead to the same result:

\begin{eqnarray}
\delta t^{(2-loop)} & = &  (g^{2} \frac{m_{H}^{2}}{m_{W}^{2}})^{2} \,
                           (\frac{m_{H}^{2}}{4 \pi \mu^{2}})^{\epsilon} \,
                           \frac{m_{H}^{2}}{(16 \pi^{2})^{2}} \, [ \,
 {{45}\over {16\,{{\epsilon }^2}}} +
 \frac{1}{\epsilon}    ( -{{33}\over 8}
     \nonumber \\
& &  + {{45\, \gamma  }\over
        {16}} + {{{9\,\sqrt{3}}\,\pi }\over {16}} ) +
  {{609}\over {128}} - {{33\, \gamma  }\over 8} +
  {{45\,{{ \gamma  }^2}}\over {32}} -
  {{45\,\sqrt{3}\,\pi }\over {64}}
     \nonumber \\
& &    +  {{{9\,\sqrt{3}}\, \gamma  \,\pi }\over
    {16}} - {{3\,{{\pi }^2}}\over {32}}
     - {{21\,\sqrt{3}\,
      {\it Cl}({{\pi }\over 3})}\over {32}} +
  {{{9\,\sqrt{3}}\,\pi \,\log (3)}\over {32}}
  \, ]
      \; \; .
\end{eqnarray}

Note that this expression disagrees with the results
of ref. \cite{maher}\footnote{At least two mistakes
exist in ref. \cite{maher}. Their diagram ${\cal L}_{20}$
has the value $\frac{\pi^{2}}{3}$, and not $\zeta(2) + 4 \, \log 2$;
and ${\cal L}_{3} = - \frac{\pi^{2}}{6} + 2 \sqrt{3} Cl(\frac{\pi}{3})$,
so it does not contain the logarithm of the golden ratio, as claimed
in those papers. These two diagrams enter the expressions of both
Goldstone and Higgs selfenergies, and hence affect all their results.}.

The main task is to calculate the two--loop diagrams of the
Higgs  selfenergy. The topologies involved are shown in fig. 5.

Subtracting the tadpole counterterm from the diagrams of fig. 5,
and performing the necessary integrations numerically,
one obtains the following Higgs selfenergy at
${\cal O}((g^{2} \frac{m_{H}^{2}}{m_{W}^{2}})^{2})$:

\begin{eqnarray}
\lefteqn{\Sigma_{HH}^{(2-loop)} (k^2=m_{M}^{2}) \, =}  \nonumber \\
& &    (g^{2} \frac{m_{H}^{2}}{m_{W}^{2}})^{2} \,
     (\frac{m_{H}^{2}}{4 \pi \mu^{2}})^{\epsilon} \,
     \frac{m_{H}^{2}}{(16 \pi^{2})^{2}} \, [ \,
 - {{9}\over {{{\epsilon }^2}}} +
 \frac{3}{32\,\epsilon }  ( 169 - 96\,\gamma -  24\,\sqrt{3}\,\pi  )
     \nonumber \\
& & \; \; \; \; \; \; - 4.6298(7) - i \, 0.4124(5)
   \, ]
      \; \; .
\end{eqnarray}

The real part of this expression gives the two--loop
Higgs mass counterterm. The absorptive part agrees with
the ${\cal O}(g^{2} \frac{m_{H}^{2}}{m_{W}^{2}})$
corrections to the Higgs decay width \cite{marciano}:

\begin{eqnarray}
\lefteqn{\Gamma (H \rightarrow W^{+}W^{-} , ZZ) \, =}  \nonumber \\
& & g^{2} \frac{m_{H}^{3}}{m_{W}^{2}} \frac{3}{128 \, \pi} \,
[ \, 1 + g^{2} \frac{m_{H}^{2}}{m_{W}^{2}} \frac{1}{8 \, \pi^{2}} \,
( \frac{19}{16} - \frac{3 \sqrt{3} \pi}{8} + \frac{5 \pi^{2}}{48} )
 \, ]
      \; \; .
\end{eqnarray}

Note that this is not only a check of the imaginary part of the diagrams.
The path in the Feynman parameters integrals was varied in large
limits to check that the result is the same. This proves that the
integrand is indeed analytical. Since an analytical function is
uniquely determined by its imaginary part, it is to be concluded that the
real part of the numerical integration is correct as well.

The momentum dependent selfenergy of the Higgs field
is given by:

\begin{eqnarray}
\lefteqn{\Sigma_{HH}^{(2-loop)} (k^2) \, =}  \nonumber \\
& &    (g^{2} \frac{m_{H}^{2}}{m_{W}^{2}})^{2} \,
     (\frac{m_{H}^{2}}{4 \pi \mu^{2}})^{\epsilon} \,
     \frac{m_{H}^{2}}{(16 \pi^{2})^{2}} \, [ \,
 - {{9}\over {{{\epsilon }^2}}} +
 \frac{3}{32\,\epsilon }  ( 168 + \frac{k^{2}}{m_{H}^{2}}
     \nonumber \\
& & \; \; \; \; \; \; - 96\,\gamma -  24\,\sqrt{3}\,\pi  )
                      + \Sigma_{f} (\frac{k^2}{m_{H}^{2}})
   \, ]
      \; \; .
\end{eqnarray}

The behaviour of its finite part $\Sigma_{f} (\frac{k^2}{m_{H}^{2}})$
is shown in fig.6. It has the expected analytical structure.
$\Sigma_{f}$, as well as the one--loop selfenergy,
displays unphysical singularities at $k^2=0$ and $k^2=4 m_{H}^{2}$,
which cancel in the full Green's functions. The imaginary part of the
selfenergy grows like $k^2$ at large momenta. Considering the asymptotic
behaviour of the phase space factor of the three--body decay, this
agrees with the Cutosky rule.

The real part of the derivative of the selfenergy at $k^2=m_{H}^{2}$
is absorbed in the wave function renormalization of the Higgs field.
On the contrary, its imaginary part is physical,
and gives the leading radiative corrections to the Higgs shape.
The momentum dependence of $Im \, \Sigma_{f} (k^2)$ describes
corrections to the Breit--Wigner shape, corresponding to an
energy dependent width.
By expanding the selfenergy around the peak and keeping
only the first derivative, one obtains the following correction
to the Higgs propagator:

\begin{eqnarray}
\frac{1}{k^{2} - m_{H}^{2} + i m_{H} \Gamma_{H}}
 & \rightarrow &
\frac{1}{k^{2} - m_{H}^{2} + i m_{H} \Gamma_{H}
               + i (k^{2} - m_{H}^{2}) \Gamma_{H}^{\prime} }
      \; \; , \nonumber
\end{eqnarray}
\begin{eqnarray}
\Gamma_{H}^{\prime} \; = \;
- (\frac{g^{2}}{16 \, \pi^{2}}
   \frac{m_{H}^{2}}{m_{W}^{2}})^{2}
\left.
Im \frac{\partial \Sigma_{f}(k^{2})}{\partial k^{2}}
\right|_{k^{2}=m_{H}^{2}}
& \approx &  1.0 \;  (\frac{g^{2}}{16 \, \pi^{2}}
                      \frac{m_{H}^{2}}{m_{W}^{2}})^{2}
      \; \; .
\end{eqnarray}

One readily sees on dimensional grounds that such an effect
does not appear at the one--loop level, because the phase space
factor is $s$ independent, and therefore no $s$--dependent
width can occur at order $g^{2}\frac{m_{H}^{2}}{m_{W}^{2}}$ .

The numerical value of $\Gamma_{H}^{\prime}$ given in eq. 35,
obtained from the momentum dependence of the Higgs
selfenergy shown in fig.6,
agrees with the exact result which can be derived
by using the Cutkosky rule:

\begin{eqnarray}
 (g^{2} \frac{m_{H}^{2}}{m_{W}^{2}})^{2} \,
  \frac{1}{(16 \pi^{2})^{2}} \,
  \frac{3 \, \pi}{4} \,
  \left(
  1 + \frac{\pi \, \sqrt{3}}{12} - \frac{5 \, \pi^{2}}{48}
  \right) & = & \nonumber \\
  = \; \;
  1.002245142 \, (g^{2} \frac{m_{H}^{2}}{m_{W}^{2}})^{2} \,
               \frac{1}{(16 \pi^{2})^{2}}
     \; \; .
\end{eqnarray}

This correction shifts the peak of the Higgs resonance towards
a lower energy, and increases its height a little, as shown
in fig.7. Further corrections, which come from the second derivative
of the Higgs selfenergy, are much smaller, and do not affect
the position of the resonance --- they only make it deviate more from
the Breit--Wigner shape.

\begin{table}
\begin{tabular}{||c||c|c|c||}                       \hline\hline
           &  peak      & peak      & height     \\
 $m_{H}$   &  shift     & shift     & increase   \\
 $[$GeV$]$ &  $[$GeV$]$ & $[$\%$]$  & $[$\%$]$   \\ \hline\hline
   400     &  -.072     & -.018     & .0020      \\ \hline
   500     &  -.35      & -.070     & .012       \\ \hline
   600     &  -1.3      & -.21      & .051       \\ \hline
   700     &  -3.8      & -.54      & .18        \\ \hline
   800     &  -9.9      & -1.2      & .51        \\ \hline
   900     &  -23       & -2.6      & 1.3        \\ \hline
   1000    &  -49       & -4.9      & 3.0        \\ \hline
   1100    &  -97       & -8.8      & 6.5        \\ \hline
   1200    &  -178      & -15       & 13         \\ \hline
   1300    &  -305      & -23       & 25         \\ \hline
   1400    &  -487      & -35       & 45         \\ \hline
   1500    &  -721      & -48       & 78         \\ \hline
   1600    &  -997      & -62       & 130        \\ \hline\hline
\end{tabular}
\caption{Leading corrections to the shape of
the Higgs resonance
$|
\frac{1}{p^{2} - m_{H}^{2} - i Im \Sigma}
|^{2}$.
The Breit--Wigner resonance with constant width calculated
at order
$(g^{2} \frac{m_{H}^{2}}{m_{W}^{2}})^{2}$ is compared
to the resonance corrected for the energy dependence
of the two--loop selfenergy.
}
\end{table}

The magnitude of these corrections for various values of
$m_{H}$ is given in table 1. The corrections become large
if the Higgs boson is heavier than $\sim$1.2 TeV, signaling
strong couplings in the electroweak symmetry breaking sector.
This agrees with well--known results based on the
unitarity violation in vector boson scattering at tree level
\cite{lee-quigg}, and on the magnitude
of one--loop radiative corrections to the Higgs width
\cite{marciano}. On the other hand, this perturbative
bound is considerably lower than the value of 3---4 TeV
derived from two--loop corrections to the
$\rho$ parameter and to the selfcouplings of the vector bosons
\cite{vdBij:2loop:rho,vdBij:2loop:vertex}. This was to
be expected, since the latter are corrections to low
energy parameters, therefore subject to the screening
theorem --- the leading contributions in $m_{H}$ cancel.


\section{Conclusions}

We described a method to calculate two--loop massive Feynman diagrams
which can be applied, at least in principle, to any diagram.

The diagrams are integrated analytically as far as possible. The
necessary formulae are easy to encode into an algebraic computer
program. The remaining integrals are made smooth
and free of singularities by appropriate changes of variables and
convenient choices of the integration paths. The resulting formulae
are suitable for numerical integration since they have a small variance.

The techniques were compared to analytical and numerical results
for certain two--loop diagrams. In all cases good agreement with
the known results was found.

The complexity of Green's functions which can be calculated
by using this method is limited only by the time needed to perform
the numerical integrations.

For less than four Feynman parameters, deterministic numerical integration
methods were used, which lead to fast and accurate answers. In particular,
two and three point functions appear to pose no problem.

For more than four Feynman parameters to integrate over,
the Monte--Carlo integration is expected to be faster, but one
has to content oneself with 3--4 digits accuracy. The time
needed to calculate the integral is not expected to grow too fast with the
number of Feynman parameters, since the speed of the Monte--Carlo
integrations depends primarily on the variance of the integrand,
and not on the dimension of the integral.

To show that the method can be used to calculate Green's
functions of physical relevance, which usually involve
many diagrams and large mass splittings,
it was used to calculate the
selfenergy of the Higgs boson and to extract the leading
corrections to the shape of the Higgs resonance.

The corrections to the Higgs shape imply a shift
of the resonance towards a lower energy. In the range
up to 800 GeV, which will be covered by the four lepton events
at LHC, the shift is quite marginal, at best at 1.2\% level.
The shift becomes increasingly important for heavier Higgs bosons,
which can presumably be searched for at LHC by looking
at the jet decay modes. However, to determine
the implications of the peak shift for the search
for a very heavy Higgs boson would request
a full analysis of the production mechanism.

The shape corrections grow with $m_{H}$, becoming rather large
for $m_{H} \sim$1.2 TeV, where the perturbative approach
eventually breaks down.

\vspace{.5cm}


\newpage


{\bf Figure captions }

\vspace{2cm}

{\em Fig.1}    The diagrammatical representation of the
               function ${\cal G}(m_{1},m_{2},m_{3};k^{2})$ (a)
	       and of its higher order derivatives (b).

\vspace{.5cm}

{\em Fig.2}    The behaviour of
               $\tilde{g}(m_{1},m_{2},m_{3};k^{2};x)$
	       as a function of $x$ under the threshold (a)
	       and above (b). The solid line represents
	       the real part, and the dashed one---the imaginary part.

\vspace{.5cm}

{\em Fig.3}    The real (a) and the imaginary (b) parts of
               $\tilde{g}(m_{1},m_{2},m_{3};k^{2};x)$
	       as a function of the complex Feynman parameter $x$.
	       The logarithmic singularities at $x=0$ and $x=1$,
	       as well as the two branching points close to the
	       real axis between 0 and 1 can be seen.
	       The integration path with the ends at 0 and 1
	       avoids the first singularity by going through
	       the positive imaginary half plane, then changes
	       to the negative imaginary half plane to avoid
	       the second singularity.

\vspace{.5cm}

{\em Fig.4}    The topologies of the two--loop tadpole diagrams.

\vspace{.5cm}

{\em Fig.5}    The topologies of the two--loop selfenergy diagrams.

\vspace{.5cm}

{\em Fig.6}    The real (solid line) and the imaginary (dashed line)
               parts of the finite part
	       $\Sigma_{f} (\frac{k^2}{m_{H}^{2}})$
	       of the selfenergy of the Higgs boson.

\vspace{.5cm}

{\em Fig.7}    The leading corrections to the shape
               of the Higgs resonance \linebreak
	       $|\frac{1}{p^{2} - m_{H}^{2} - i Im \Sigma}|^{2}$.
	       The Breit--Wigner resonance with constant width
	       (solid line), and the correction for the
	       energy dependence of the two--loop self--energy
	       (dashed line).

\end{document}